\documentclass[conference,a4paper]{IEEEtran}
\IEEEoverridecommandlockouts

\usepackage[draft]{graphicx}
\usepackage{balance}
\usepackage[T1]{fontenc}
\usepackage{ifthen}
\usepackage[cmex10]{amsmath} % Use the [cmex10] option to ensure
\usepackage[lined,boxed,commentsnumbered,linesnumbered, ruled]{algorithm2e}
\usepackage[url,hyperrefblack,notheorems,IEEEtran]{research17}
\hypersetup{draft}
\usepackage{lipsum} % print dummy text
\usepackage{comment}
\usepackage{booktabs} % required for showing good \hline in matrix form --> \midrule
\usepackage{mathtools}
\usepackage{amsmath,amssymb,amsfonts,amsthm}
\usepackage{tikz} % common graphical tools for pictures
\usetikzlibrary{shapes, patterns,decorations.text, decorations.pathreplacing}

\newtheorem{theorem}{\mytheoremname}
\newtheorem{definition}{\mydefinitionname}

\newtheorem{corollary}{\mycorollaryname}

\newtheorem{remark}{\myremarkname}

\usepackage[capitalize]{cleveref}
\crefname{equation}{\unskip}{\unskip}
\crefname{claim}{Claim}{Claims} %
\usepackage{array}
\usepackage{multirow}
\newcolumntype{C}[1]{>{\centering\arraybackslash}p{#1}}
\renewcommand{\vect}[1]{\vectg{#1}} % switch by default to second version!
\newcommand*{\Resize}[2][4]{\resizebox{#1}{!}{\ensuremath{#2}}} % Resize based on line or column width
\makeatletter
\renewcommand*\env@matrix[1][*\c@MaxMatrixCols c]{%
  \hskip -\arraycolsep
  \let\@ifnextchar\new@ifnextchar
  \array{#1}}
\makeatother
\newcommand{\HP}[1]{\HH\left(#1\right)} 
\newcommand{\eHP}[1]{\HH(#1)} 
\newcommand{\bigHP}[1]{\HH\bigl(#1\bigr)}

\newcommand{\HPcond}[2]{\HH\left(#1 \kern0.1em\middle|\kern0.1em #2\right)}
\newcommand{\eHPcond}[2]{\HH(#1 \kern0.1em|\kern0.1em #2)} 
\newcommand{\bigHPcond}[2]{\HH\bigl(#1 \kern-0.1em \bigm| \kern-0.1em#2\bigr)}
\newcommand{\BigHPcond}[2]{\HH\Bigl(#1 \kern-0.1em \Bigm| \kern-0.1em#2\Bigr)}
\newcommand{\MI}[2]{\II\left(#1 \kern0.1em{;}\kern0.1em #2\right)} 
\newcommand{\eMI}[2]{\II(#1 \kern0.1em{;}\kern0.1em #2)} 
\newcommand{\bigMI}[2]{\II\bigl(#1 \kern0.1em{;}\kern0.1em #2\bigr)}
\newcommand{\BigMI}[2]{\II\Bigl(#1 \kern0.1em{;}\kern0.1em #2\Bigr)}
\newcommand{\MIcond}[3]{\II\left(#1 \kern0.1em{;}\kern0.1em #2 \kern0.1em\middle|\kern0.1em #3\right)}
\newcommand{\eMIcond}[3]{\II(#1 \kern0.1em{;}\kern0.1em #2 \kern0.1em|\kern0.1em #3)} 
\newcommand{\bigMIcond}[3]{\II\bigl(#1 \kern0.1em{;}\kern0.1em #2 \kern-0.1em \bigm| \kern-0.1em#3\bigr)}
\newcommand{\BigMIcond}[3]{\II\Bigl(#1 \kern0.1em{;}\kern0.1em #2 \kern-0.1em \Bigm| \kern-0.1em#3\Bigr)}
\DeclareSymbolFont{matha}{OML}{txmi}{m}{it}% txfonts
\DeclareMathSymbol{\varv}{\mathord}{matha}{118}
\usepackage{upgreek}

\begin{document}
\sloppy \title{On the Capacity of Private Nonlinear  Computation for Replicated Databases \thanks{This work is supported by US NSF grant CNS-1526547.}}

\author{
		
		\IEEEauthorblockN{Sarah A.~Obead\IEEEauthorrefmark{2}, Hsuan-Yin Lin\IEEEauthorrefmark{3}, Eirik
		Rosnes\IEEEauthorrefmark{3}, and J{\"o}rg Kliewer\IEEEauthorrefmark{2} \\
		\IEEEauthorblockA{\IEEEauthorrefmark{2}Helen and John C.~Hartmann
			Department of Electrical and Computer Engineering \\ New Jersey Institute of Technology, Newark, New Jersey 07102, USA\\}
		\IEEEauthorblockA{\IEEEauthorrefmark{3}Simula UiB, N--5006 Bergen, Norway}}}

\maketitle

\begin{abstract}
We consider the problem of private computation (PC) in a distributed storage system. In such a setting a user wishes to
	compute a function of $f$ messages replicated across $n$ noncolluding databases, while revealing no information about the
	 desired function to the databases. We provide an information-theoretically accurate achievable PC rate, which is the ratio of the smallest desired amount of information and the total amount of downloaded information, for the scenario of nonlinear computation. For a large message size the rate equals the PC capacity, i.e., the maximum achievable PC rate, when the candidate functions are the $f$ independent messages and one arbitrary nonlinear function of these. When  the number of messages grows, the
    PC rate approaches an outer bound on the PC capacity. %
	 As a special case, we consider private monomial computation (PMC)  %
	 and numerically compare the achievable PMC rate to the outer bound for a finite number of messages.
\end{abstract}


\section{Introduction}
\label{sec:introduction}

The problem of private information retrieval (PIR) from public databases,
introduced in 
\cite{ChorGoldreichKushilevitzSudan95_1}, has been the focus of attention for several decades in the
computer science community (see, e.g., \cite{Gasarch04_1,Yekhanin10_1}). 
In PIR, the goal is to privately access an arbitrary
message stored in a database without revealing any information of the identity of the desired message. If the users do not have any side information on the data stored in the database, the best strategy is to store the
messages in at least two databases while ensuring PIR. Hence,
the design of PIR protocols has focused on the case when multiple databases, i.e., distributed storage systems (DSSs), store the messages. 
Recently, the aspect of minimizing the communication cost, e.g., the required rate or bandwidth of privately querying the databases with the
desired requests and downloading the corresponding information  has attracted a great deal of attention in the information theory and coding communities. Thus, the renewed interest in PIR primarily focused on the study and design of efficient PIR protocols for DSSs. 
For example, \cite{SunJafar17_1,SunJafar18_2}, presented fundamental limits of the PIR rate when data is replicated over noncolluding and colluding databases, respectively. 

Motivated by privacy concerns in distributed computing, a generalization of the PIR problem has emerged recently
\cite{SunJafar19_2,MirmohseniMaddahAli18_1,ChenWangJafar18_1, ObeadKliewer18_1,ObeadLinRosnesKliewer18_1,Karpuk18_1,RavivKarpuk19_1app, ObeadLinRosnesKliewer19_1app} to address the
\emph{private computation (PC)} of arbitrary functions over the stored messages. 
In PC a user intends to compute a function of the messages  stored at multiple databases while keeping the identity of the function private from each database, as they may be under the control of an adversary. In
\cite{SunJafar19_2,MirmohseniMaddahAli18_1}, the scenario of private linear computation (PLC) is considered for noncolluding replicated databases. In these works, the capacity and achievable rates for the communication
overhead needed to privately compute a given \emph{linear} function
were derived as a function of the number of messages and the number of
databases, respectively. Interestingly, the PLC capacity is equal to the PIR capacity of
\cite{SunJafar17_1}. 
The extension to the coded case is addressed in \cite{ObeadKliewer18_1,ObeadLinRosnesKliewer18_1} and \cite{Karpuk18_1,RavivKarpuk19_1app,ObeadLinRosnesKliewer19_1app} for PLC  and private polynomial  computation (PPC), %
respectively. 

In contrast to our previous work in \cite{ObeadLinRosnesKliewer19_1app} (and
also \cite{Karpuk18_1,RavivKarpuk19_1app}), which considered 
PPC schemes for coded storage for polynomials of degree at most $g$, for some fixed integer $g$,  and only  a simplified rate definition, in this work
we extend these considerations to general private nonlinear computation for replication-based
storage and an \emph{exact} information-theoretic definition of the PC rate.  This
complicates the analysis. We also include a converse result which is absent
from \cite{ObeadLinRosnesKliewer19_1app}. We provide a general achievable scheme for the scenario of nonlinear
computation with rate equal to the PC capacity, i.e., the maximum achievable PC rate, when the message size is large and
the candidate functions are the independent messages and one arbitrary nonlinear function of these. When the number of messages grows, the PC rate approaches an outer bound on the PC capacity derived from \cite[Thm.~1]{ChenWangJafar18_1} and thus becomes the capacity itself. A similar result was stated in \cite[Thm.~2]{SunJafar19_2}, however for a simplified definition of the PC rate 
that does not take into account that the candidate functions may have different amount of information, referred to as the function size.  
Moreover, we discuss how a 
PC scheme should be designed to achieve the PC capacity. As a special case, we consider private monomial
computation (PMC) and numerically compare the achievable PMC rate to the outer bound for a finite number of messages.

%%%%%%%%%%%%%%%%%%%%%%%%%%%%%%%%%%%%%%%%%%%%%%%%%%%%%%%%%%%%%%%%%%%%%%%%%%%%%%%%%%%%%%%%%%%%%%%%%%%%%%%%%%%%%%%%%%%%%%%%%
\section{Preliminaries}
\label{sec:preliminaries}

\subsection{Notation}
\label{sec:notation}

We denote by $\Naturals$ the set of all positive integers, $[a]\eqdef\{1,2,\ldots,a\}$, and
$[a:b]\eqdef\{a,a+1,\ldots,b\}$ for $a,b\in \Naturals$, $a \leq b$.  
Random and deterministic quantities are carefully distinguished as follows.
A random variable is denoted by a capital Roman letter, e.g., $X$, while its realization is denoted by the corresponding small
Roman letter, e.g., $x$. Vectors are boldfaced, e.g., $\vect{X}$ denotes a random vector and $\vect{x}$ denotes a
deterministic vector.  
In addition, sets are denoted by calligraphic uppercase letters, e.g., $\set{X}$. The notation $\vect{X} \sim \vect{Y}$ is used to indicate that $\vect{X}$ and $\vect{Y}$ are identically distributed. 
For a given index set $\set{S}$, we also
write $\vect{X}^\set{S}$ to represent $\bigl\{\vect{X}^{(v)}\colon v\in\set{S}\bigr\}$. Furthermore, some constants and functions are also depicted by
Greek letters or a special font, e.g., $\const{X}$.
The function $\HP{X}$ represents the entropy of $X$, and $\MI{X}{Y}$ the mutual
information between the random variables $X$ and $Y$. %
The binomial coefficient of $a$ over $b$ is denoted by $a \choose b$. %

A monomial ${\vect{z}}^{\vect{i}}$ in $m$ variables $z_1,\ldots,z_m$ with degree $g$ is written as
${\vect{z}}^{\vect{i}} = z_1^{i_1} \cdots z_m^{i_m}$, where
$\vect{i}\eqdef(i_1,\ldots,i_m)\in (\{0\} \cup \Naturals)^{m}$ is the exponent vector with
$\mathsf{wt}(\vect{i})\eqdef\sum_{j=1}^{m}i_j = g$. The set
$\{\vect{z}^{\vect{i}}: \vect{i} \in (\{ 0 \} \cup \Naturals)^m,\, 1\leq \mathsf{wt}(\vect{i})\leq g\}$ of all monomials
in $m$ variables of degree at most $g$ has size
\begin{IEEEeqnarray*}{rCl}
  \const{M}(m,g)\eqdef\sum\limits_{h=1}^{g}\binom{h+m-1}{h}=\binom{g+m}{g}-1.
\end{IEEEeqnarray*}

\subsection{Problem Statement}
\label{sec:rep-PMCsystem-model}

The PC problem for replicated DSSs is described as follows. We consider a DSS that stores in total $f$ independent
messages $\vect{W}^{(1)},\ldots,\vect{W}^{(f)}$, where each message $\vect{W}^{(m)}=\bigl(W_{1}^{(m)},\dots,W_{\beta\const{L}}^{(m)}\bigr)$, $m\in [f]$, is a random length-$\beta\const{L}$ vector with independent and identically distributed symbols that are chosen at random from the field $\Field_q$ for some
$\beta,\const{L} \in\Naturals$. The messages are replicated and stored in the $j$-th database, $j\in[n]$.  Without loss of generality, we assume that the symbols of each message are selected uniformly over the field  $\Field_q$. Thus,
\begin{IEEEeqnarray*}{rCl}
  \bigHP{\vect{W}^{(m)}}& = &\beta\const{L},\,\forall \,m\in[f],
  \\
  \bigHP{\vect{W}^{(1)},\dots,\vect{W}^{(f)}}& = &f\beta\const{L}\quad (\textnormal{in } q\textnormal{-ary units}).
\end{IEEEeqnarray*}

We consider the case of $n$ noncolluding databases. In PC, a user wishes to
privately compute exactly one function image $X_i^{(v)} \triangleq \phi^{(v)}(W_i^{(1)},\ldots,W_i^{(f)})$, $\forall\, i\in [\beta\const{L}]$, out of $\mu$ arbitrary \emph{candidate} functions $\phi^{(1)},\ldots,\phi^{(\mu)}{\colon(\Field_q)^f\to\Field_q}$, where %
$X^{(v)}_1,\ldots,X^{(v)}_{\beta\const{L}}$ are independent and identically distributed according to a prototype random variable $X^{(v)}$ with probability mass function $P_{X^{(v)}}$. 
Now, let $\vect{X}^{(v)} \triangleq \bigl(X^{(v)}_1,\ldots,X^{(v)}_{\beta\const{L}}\bigr)$. With some abuse of language, in the following, we often refer to the image $\vect{X}^{(v)}$ as the function $\phi^{(v)}$. Without loss of generality, we assume that the
candidate functions are ordered descendingly with respect to their entropy,
i.e.,~$\HP{X^{(1)}}=\max_{v\in[\mu]}{\HP{X^{(v)}}}\eqdef\HH_\textnormal{max}$ and
$\HP{X^{(\mu)}}=\min_{v\in[\mu]}{\HP{X^{(v)}}}\eqdef\HH_\textnormal{min}$. Thus, in $q$-ary units, we have
\begin{IEEEeqnarray*}{rCl}
  \eHP{\vect{X}^{(v)}}& = &
  \beta\const{L}\bigHP{X^{(v)}},\,\forall\,v\in[\mu], 
  \\
  \eHP{\vect{X}^{(1)},\ldots,\vect{X}^{(\mu)}}& = &
  \beta\const{L}\HP{X^{(1)},\ldots,X^{(\mu)}},
  \\
  \eHP{{X}^{(1)}}\geq\eHP{X^{(2)}}& \geq &\dots \geq \eHP{X^{(\mu)}} \geq 0.
\end{IEEEeqnarray*} 
The user privately selects an index $v\in[\mu]$ and wishes to compute the $v$-th function while
keeping the requested function index $v$ private from each database. 
In order to retrieve the desired function $\vect{X}^{(v)}$, $v\in[\mu]$, from the DSS,  the user sends a random query $Q^{(v)}_j$
 to the $j$-th database for all $j\in[n]$. The queries are generated by the user without any prior
knowledge of the realizations of the stored messages, and they are independent of the candidate functions. In other
words, %
 $\MI{\vect{X}^{(1)},\ldots,\vect{X}^{(\mu)}}{Q^{(v)}_1,\ldots,Q^{(v)}_n}=0$, $\forall\,v\in[\mu]$.

In response to the received query, the $j$-th database sends the answer $A^{(v)}_j$ back to the user, where  $A^{(v)}_j$ is a
deterministic function of $Q^{(v)}_j$ and the data stored in the database. Thus, 
 $\bigHPcond{A^{(v)}_j}{Q^{(v)}_j,\vect{W}^{[f]}}=0$, $\forall\,v\in[\mu]$ and $\forall\,j\in[n]$.
 
 To maintain user privacy, the query-answer function must be identically distributed for all possible function indices
 $v\in[\mu]$ from the perspective of each database. In other words, the scheme's queries and answer
 strings must be independent from the desired function index. Moreover, the user must be able to reliably decode the
 desired function $\vect{X}^{(v)}$ from the received database answers.

  Consider a DSS with $n$ noncolluding replicated databases storing $f$ messages. The user wishes to retrieve the $v$-th
  function $\vect{X}^{(v)}$, $v\in[\mu]$, from the queries $Q^{(v)}_j$ and answers $A^{(v)}_j$, $j\in[n]$. For a PC
  protocol, the following conditions must be satisfied $\forall\,v,v' \in[\mu]$, $v \neq v'$, and $\forall\,j\in[n]$, 
  \begin{IEEEeqnarray*}{rCl}    
    &&\textnormal{[Privacy]} \\%
    && \qquad\quad (Q^{(v)}_j,A^{(v)}_j,\vect{X}^{[\mu]})  \sim (Q^{(v')}_j,A^{(v')}_j,\vect{X}^{[\mu]}),
    \\
    &&\textnormal{[Recovery]} \\%
    && \qquad\quad\bigHPcond{\vect{X}^{(v)}}{A^{(v)}_1,\ldots,A^{(v)}_n,Q^{(v)}_1,\ldots,Q^{(v)}_n}=o(\const{L}),
  \end{IEEEeqnarray*}
 where any function of $\const{L}$, say $\lambda(\const{L})$, is said to be $o(\const{L})$ if $\lim_{\const{L} \to \infty} \lambda(\const{L})/\const{L} = 0$.

To measure the efficiency of a PC protocol, we consider the required number of downloaded symbols for retrieving the
$\beta\const{L}$ symbols of the desired function.

\begin{definition}[PC rate and capacity for replicated DSSs]
  \label{def:def_PCrate}
  The rate of a PC protocol, denoted by $\const{R}$, is defined as the ratio of the smallest desired function size
  $\beta \const{L}\HH_\textnormal{min}$ to the total required download cost $\const{D}$, i.e.,\footnote{We adopt the rate definition of
    the dependent PIR (DPIR) problem \cite{ChenWangJafar18_1}.}
  \begin{IEEEeqnarray*}{c}
    \const{R}\eqdef\frac{\beta\const{L}\HH_\textnormal{min}}{\const{D}}.
  \end{IEEEeqnarray*}
  The PC \emph{capacity}, denoted by $\const{C}_\textnormal{PC}$, is the maximum achievable PC rate over all possible PC protocols.
\end{definition}

%%%%%%%%%%%%%%%%%%%%%%%%%%%%%%%%%%%%%%%%%%%%%%%%%%%%%%%%%%%%%%%%%%%%%%%%%%%%%%%%%%%%%%%%%%%%%%%%%%%%%%%%%%%%%%%%%%%%%%%%%%%%%%%%%%%%%%%%%%
\section{A Converse Bound and an Achievable Scheme}
\label{sec:PCachievable-rate}

In this section, we first derive an outer bound on the PC rate of any PC protocol from  \cite[Thm.~1]{ChenWangJafar18_1} (Theorem~\ref{th:converse_bound} below) and then an achievable rate for the special case of large message sizes (Theorem~\ref{thm:PCachievable-rate} below).

\subsection{Converse Bound}
\begin{theorem} \label{th:converse_bound}
Consider a DSS with $n$ noncolluding replicated databases storing $f$ messages, where the number of arbitrary candidate functions to be computed  is $\mu \geq 1$. Then, the PC capacity $\const{C}_\textnormal{PC}$ is upperbounded as
\begin{IEEEeqnarray}{rCl}
 \const{C}_\textnormal{PC}& \leq & \frac{n^{\mu}\HH_\textnormal{min}}{\sum\limits_{v=1}^\mu n^{\mu-v+1}\bigl[\eHP{X^{[v]}}-\eHP{X^{[v-1]}}\bigr]},
  \label{eq:R_upperbound}
  \end{IEEEeqnarray}
  where $X^{[0]}$ is the empty set  and $\eHP{\emptyset}=0$.  %equal to zero.
\end{theorem}

\begin{IEEEproof}
From the converse proof of either \cite{SunJafar19_2} or \cite{ChenWangJafar18_1}, it is not difficult to see that
the total download cost $\const{D}$ of a PC protocol is lowerbounded as
	\begin{IEEEeqnarray*}{rCl}
	\const{D}& \geq &  \bigHP{\vect{X}^{(1)}}+\frac{\bigHPcond{\vect{X}^{(2)}}{\vect{X}^{(1)}}}{n}
	+\frac{\bigHPcond{\vect{X}^{(3)}}{\vect{X}^{(1)},\vect{X}^{(2)}}}{n^2}
	\nonumber\\[2mm]
	&&\qquad\>+\cdots+\frac{1}{n^{\mu-1}}\bigHPcond{\vect{X}^{(\mu)}}{\vect{X}^{(1)},\ldots,\vect{X}^{(\mu-1)}},
	\IEEEeqnarraynumspace\label{eq:lowerbound_D}
\end{IEEEeqnarray*}
from which the result follows directly from Definition~\ref{def:def_PCrate}. 
\end{IEEEproof}

\begin{corollary}
  \label{cor:Outerbound}
  The outer bound from \eqref{eq:R_upperbound} equals 
    \begin{IEEEeqnarray}{c}
		\HH_\textnormal{min} \frac{1-\frac{1}{n}}{1-{(\frac{1}{n})}^f}\eqdef\HH_\textnormal{min}\const{C}_\textnormal{PIR}
		\label{eq:PMCcapacity}
	\end{IEEEeqnarray}
	when  $\mu\geq f$ and the candidate functions include the $f$ independent messages $\vect{W}^{(1)},\ldots,\vect{W}^{(f)}$, 
	where $\const{C}_\textnormal{PIR} =  \frac{1-\frac{1}{n}}{1-{(\frac{1}{n})}^f}$ is the PIR capacity for a DSS with $n$ noncolluding replicated databases storing $f$ messages \cite{SunJafar17_1}. %
 \end{corollary}

\subsection{Achievability}

\begin{theorem}
  \label{thm:PCachievable-rate}
  Consider a DSS with $n$ noncolluding replicated databases storing $f$ messages of length $\beta \const{L}$, where the number of arbitrary candidate functions to be computed  is $\mu \geq 1$.
  Then, as $\const{L} \to \infty$, the PC rate
  \begin{IEEEeqnarray}{c}
    \const{R}=\frac{\HH_\textnormal{min}}{\sum\limits_{v=1}^{\mu-1}\frac{1}{n^{v-1}}\eHP{X^{(v)}}+
      \frac{1}{n^{\mu-1}}\Bigl[\eHP{X^{[\mu]}}-\sum\limits_{v=1}^{\mu-1}\eHP{X^{(v)}}\Bigr]}
    \label{eq:PCachievable-rate}\IEEEeqnarraynumspace
  \end{IEEEeqnarray}
  is achievable.
\end{theorem}

\begin{corollary}
  \label{cor:lowerbound_R}
  The PC rate $\const{R}$ from \eqref{eq:PCachievable-rate} is lowerbounded as
  \begin{IEEEeqnarray*}{c}
    \const{R}\geq
    \frac{\HH_{\min}}{\HH_{\max}}  \frac{1-\frac{1}{n}}{1-{(\frac{1}{n})}^{\mu}}.
  \end{IEEEeqnarray*}    
\end{corollary}

\begin{corollary}
  \label{cor:R_PCbasis}
  Consider a DSS with $n$ noncolluding replicated databases storing $f$ messages of length $\beta \const{L}$. Then, as $\const{L} \to \infty$, the PC rate

  \begin{IEEEeqnarray}{rCl}
    \const{R}& = &
    \begin{cases}
     \HH_\textnormal{min}\frac{1-\frac{1}{n}}{1-\bigl(\frac{1}{n}\bigr)^f}=\HH_\textnormal{min} \const{C}_\textnormal{PIR},
      \\*\hspace*{4.5cm} \textnormal{if }\mu=f+1,
      \\[2mm]
      \frac{\HH_\textnormal{min}(1-\frac{1}{n})}{1-\bigl(\frac{1}{n}\bigr)^f %
    +\bigl(1-\frac{1}{n}\bigr)\sum\limits_{v=f+1}^{\mu-1}\eHP{X^{(v)}}\bigl[\frac{1}{n^{v-1}}-\frac{1}{n^{\mu-1}}\bigr]},
      \\*\hspace*{4.5cm} \textnormal{if }
      \mu\geq f+2
    \end{cases}
    \label{eq:R_PCbasis}
  \end{IEEEeqnarray}    
  is achievable when the candidate functions include the $f$ independent messages $\vect{W}^{(1)},\ldots,\vect{W}^{(f)}$.
\end{corollary}

\begin{remark}\leavevmode                 
  \label{rem:remark1}
  \begin{itemize}
  \item For $\mu=f+1$ the PC rate from Corollary~\ref{cor:R_PCbasis} equals the outer bound from Corollary~\ref{cor:Outerbound}. Thus, the proposed scheme is capacity-achieving.
  \item The PC rate from Corollary~\ref{cor:R_PCbasis} and the outer bound from Corollary~\ref{cor:Outerbound} converge to  $\HH_\textnormal{min}(1-1/n)$ as $f \to \infty$. A similar result was stated in \cite[Thm.~2]{SunJafar19_2}, however for a simplified definition of the PC rate.
  \item The rate of \eqref{eq:PCachievable-rate} extends the elementary capacity result for the case of two arbitrary 
    correlated functions \cite[Sec.~VII]{SunJafar19_2}, while the lower bound from Corollary~\ref{cor:lowerbound_R} matches the lower bound on the capacity of DPIR \cite[Sec.~III-B]{ChenWangJafar18_1}.
  \item If all the $\mu$ functions are uniformly distributed, ${\HH_\textnormal{min}}={\HH_\textnormal{max}}$
	and we obtain the PC rate 
    \begin{IEEEeqnarray*}{rCl}
       \const{R}& = & \frac{1-\frac{1}{n}}{1-{(\frac{1}{n})}^{\mu}}.
    \end{IEEEeqnarray*}
  \end{itemize}  
\end{remark}

A PMC problem is a PC problem where the candidate functions to be computed are restricted
to a subset of all possible multivariate monomials in $f$ variables (or messages) with degree at most $g$ which includes $\vect{W}^{(1)},\ldots,\vect{W}^{(f)}$, where $f \leq \mu \leq \const{M}(f,g)$,
$g\in\Naturals$. 
The goal here is to find a scheme that achieves the outer bound in \eqref{eq:PMCcapacity}.
Towards this goal, we state the following remark.

\begin{remark}\leavevmode
	\label{rem:remark2}
	\begin{itemize}
		\item For multivariate monomials in $f$ variables with degree at most $g$, it can be seen that the PMC rate
		\begin{IEEEeqnarray}{c}
			\frac{1-\frac{1}{n}}{1-{(\frac{1}{n})}^{\mu}}  
			\label{eq:pmc_rate}
		\end{IEEEeqnarray}
		can be achieved via the PIR protocol from \cite{SunJafar17_1} by considering each candidate monomial as a \emph{virtual} message. 
		\item In the case of monomials with degree at most $g=1$, $\mu = f$ (since ${\const{M}(f,g)}=f$) and ${\HH_\textnormal{min}}={\HH_\textnormal{max}}$, and the PMC rate reduces to the PIR capacity $\const{C}_{\textnormal{PIR}}$.
		\item Finally, for monomials with higher degree, i.e., $g \geq2$, we can achieve a PMC rate $\const{R}$ strictly larger than \eqref{eq:pmc_rate}  by Corollary~\ref{cor:R_PCbasis}, using a   
		similar approach of redundancy elimination as in the schemes in  \cite[Sec.~III-C]{ObeadLinRosnesKliewer19_1app}. Moreover, the gap between the achievable PMC rate and the outer bound from \eqref{eq:PMCcapacity} decreases with the degree of the monomials and the number of messages (see Section~\ref{sec:PMC_Redundancy}).
	\end{itemize}
\end{remark}

%%%%%%%%%%%%%%%%%%%%%%%%%%%%%%%%%%%%%%%%%%%%%%%%%%%%%%%%%%%%%%%%%%%%%%%%%%%%%%%%%%%%%%%%%%%%%%%%%%%%%%%%%%%%%%%%%%%%%%%%%
\subsection{Achievable Scheme for Theorem~\ref{thm:PCachievable-rate}}
\label{sec:achievable-scheme_uncodedDSSs}

We start with a PIR query scheme for $\mu$ %
\emph{virtual} messages, where the $\mu$ arbitrary candidate
functions of the PC problem are considered as $\mu$ arbitrary correlated messages. Given that $\mu$ virtual messages are replicated over $n$ noncolluding databases, we require the length of each message to be
$\beta\const{L}=n^{\mu}\const{L}$ with a sufficiently large $\const{L}$. Let $\vect{X}^{(v)}=(\vect{X}^{(v)}_{1},\dots,\vect{X}^{(v)}_{\beta})$, where each segment $\vect{X}^{(v)}_i$, $i\in[\beta]$, contains $\const{L}$ symbols.
For $\tau\in[\mu]$, a sum $\vect{X}_{i_1}^{(v_1)} + \cdots + \vect{X}^{(v_\tau)}_{i_\tau}$ of
$\tau$ distinct candidate function segments is called a $\tau$-sum for any $(i_1,\ldots,i_{\tau})\in [\beta]^\tau$, and
$\{v_1,\ldots,v_{\tau}\}\subseteq [\mu]$ determines the type of the $\tau$-sum.

Here, we rely on lossless data compression of large-enough message segments to achieve the PC rate presented in
Theorem~\ref{thm:PCachievable-rate}. However, due to possible dependency across message symbols associated with
the same subindex, we follow similar  index assignment and message symmetry principles as for the PLC schemes in \cite{SunJafar19_2,ObeadKliewer18_1,ObeadLinRosnesKliewer18_1}. 

The overall protocol is composed of $\mu$ rounds. For a desired function indexed by $v\in[\mu]$, a query set $Q^{(v)}_j$, $j\in[n]$, is composed of $\mu$
disjoint subsets, one generated by each round $\tau\in[\mu]$. For each round $\tau$ the query subset is further subdivided
into two subsets. The first subset $Q^{(v)}_j(\set{D};\tau)$ consists of $\tau$-sums with a single symbol from the
\emph{desired} message and $\tau-1$ symbols from \emph{undesired} messages, while the second subset $Q^{(v)}_j(\set{U};\tau)$ contains $\tau$-sums with
symbols only from undesired messages.\footnote{With some abuse of notation, the generated queries are sets containing their answers.} We let $\pi$ be a random permutation over the $\beta$ message segments. For $v\in[\mu]$,
 \begin{IEEEeqnarray*}{c}
 \vect{U}^{(v)}_{t}\triangleq {\vect{X}}^{(v)}_{\pi(t)},\quad t\in[\beta],
 \end{IEEEeqnarray*} 
denotes a permuted segment from the virtual message $\vect{X}^{(v)}$, where the permutation $\pi$ is selected privately by the
user and is applied  as a one-time pad to all messages. %
Without loss of generality, let the desired virtual message be $\vect{X}^{(1)}$. The
construction of the queries for arbitrary $n$ and $\mu$ is done round-wise for each round $\tau\in[\mu]$ and each
database as shown in Table~\ref{tab:DPIR-table}. The answer string of each database is generated as follows.

\begin{table}[tbp!]
  \centering
  \caption{Query sets for a DSS with $n$ noncolluding replicated databases storing $f$ messages and where the first ($v=1$) out of $\mu$ candidate functions is privately computed.  %
  For  simplicity, $\vect{U}^{(v)}_{\ast}$ indicates that the exact requested subindex $t\in[\beta]$ is omitted. }
    \vspace{-1.5ex}
  \label{tab:DPIR-table}
   \Resize[\columnwidth]{
    \begin{IEEEeqnarraybox}[
      \IEEEeqnarraystrutmode
      \IEEEeqnarraystrutsizeadd{4pt}{3pt}]{v/c/v/c/v/c/v/c/v} 
      \hline
      \IEEEeqnarrayrulerow \\
      %& j\textnormal{-th database} && 1  && \;\dots\; && n\\
      & j && 1  && \;\dots\; && n\\
      \hline \hline
      & Q^{(1)}_j(\set{D};1)
      && \vect{U}^{(1)}_{1}  &&  \dots && \vect{U}^{(1)}_{n} &
      \\*\cline{1-9}      
      & Q^{(1)}_j(\set{U};1)
      && \vect{U}^{(2)}_1,\ldots,\vect{U}^{(\mu)}_{1} &&  \cdots && \vect{U}^{(2)}_{n},\dots,\vect{U}^{(\mu)}_{n} &
      \\*\cline{1-9}      
      & \multirow{5}{*}{$Q^{(1)}_j(\set{D};2)$}
         && \vect{U}^{(1)}_{n+1}+\vect{U}^{(2)}_{2}  && \cdots && \vect{U}^{(1)}_{n+(\mu-1)(n-1)^2+1}+\vect{U}^{(2)}_{1} &\\
      &  && \vdots &&  \vdots && \vdots &\\
      &  && \vect{U}^{(1)}_{n+\mu-1}+\vect{U}^{(\mu)}_{2}  && \cdots && \vect{U}^{(1)}_{n+(\mu-1)(n-1)^2+(\mu-1)}+\vect{U}^{(\mu)}_{1}
      &
      \\
      &  && \vdots &&  \vdots && \vdots &\\
       &  && \vect{U}^{(1)}_{n+(\mu-1)(n-1)}+\vect{U}^{(\mu)}_{n} && \cdots && \vect{U}^{(1)}_{n+n(\mu-1)(n-1)}+\vect{U}^{(\mu)}_{n-1} &
      \\*\cline{1-9}      
      &  \multirow{3}{*}{$Q^{(1)}_j(\set{U};2)$}
         && \vect{U}^{(2)}_{n+2}+\vect{U}^{(3)}_{n+1}  && \cdots && \vect{U}^{(2)}_{*}+\vect{U}^{(3)}_{n+(\mu-1)(n-1)^2+1} &\\
         &  && \vdots && \vdots && \vdots &
         \\
      &  && \vect{U}^{(\mu-1)}_{n+(\mu-1)(n-1)}+\vect{U}^{(\mu)}_{*} && \cdots &&
      \vect{U}^{(\mu-1)}_{n+n(\mu-1)(n-1)}+\vect{U}^{(\mu)}_{*} &
      \\*\cline{1-9}      
      & \vdots
      && \;\; \vdots &&  \;\; \vdots \;\; && \;\; \vdots \;\; &
      \\*\cline{1-9}      
      &  \multirow{3}{*}{$Q^{(1)}_j(\set{D};\mu)$}
      && \vect{U}^{(1)}_{*}+\cdots+\vect{U}^{(\mu)}_{*} && \cdots &&\vect{U}^{(1)}_{*}+\cdots+\vect{U}^{(\mu)}_{*} &
      \\ 
      &  && \vdots && \vdots && \vdots &
      \\
      &  && \vect{U}^{(1)}_{*}+\cdots+\vect{U}^{(\mu)}_{*} &&\cdots &&\vect{U}^{(1)}_{n^\mu}+\cdots+\vect{U}^{(\mu)}_{*}
      &
      \\ 
      \IEEEeqnarrayrulerow
    \end{IEEEeqnarraybox}
}
  \vspace{-4ex}
\end{table}

\begin{itemize}
\item For the first round ($\tau=1$), optimally compress the length-$\const{L}$  segments
  $\bigl\{\vect{U}^{(1)}_{t},\vect{U}^{(2)}_{t},\dots,\vect{U}^{(\mu)}_{t}\bigr\}$,  $t\in[\beta]$, jointly, which results in $\const{L}
  \eHP{X^{[\mu]}} + {o}(\const{L})$ units.
\item In the second round ($\tau=2$), for the $2$-sum 
  $\vect{U}^{(v)}_{t}+\vect{U}^{(v')}_{t'}$, $\forall\,v,v'\in[\mu]$, $v < v'$, and $t,t'\in[\beta]$, compress each
  message segment independently based on ${\max}\{\eHP{X^{(v)}}, \eHP{X^{(v')}}\}$ and then return the sum of the two
  compressed segments, which results in $\const{L}\max\{\eHP{X^{(v)}},\eHP{X^{(v')}}\}+o{(\const{L})}$ units. For this
  round, one can show that in total $(n-1)\sum_{v=1}^{\mu-1} (\mu-v)\const{L}\eHP{X^{(v)}}+ o{(\const{L})}$ units are downloaded.
\item For the following rounds ($\tau>2$), each database compresses the segments of each queried $\tau$-sum 
  $\sum_{l=1}^{\tau} \vect{U}^{(v_l)}_{t_l}$, where $\{v_1,\ldots,v_\tau\} \subseteq [\mu]$ %
  and $(t_1,\ldots,t_\tau) \in [\beta]^\tau$, 
  separately based on $\max\{\eHP{X^{(v_1)}},\ldots,\eHP{X^{(v_\tau)}}\}$. Each database then returns the sum of the
  compressed segments in $\const{L}\max\{\eHP{X^{(v_1)}},\ldots,\eHP{X^{(v_\tau)}}\}+o(\const{L})$ units. By the end of
  each round, one can show that in total
  $(n-1)^{\tau-1}\sum_{v=1}^{\mu-(\tau-1)}{\mu-v\choose\tau-1}\const{L}\eHP{X^{(v)}}+o{(\const{L})}$ units are downloaded for each
  $\tau \in[3:\mu]$.
\end{itemize}

\subsubsection{Recovery and Privacy}
\label{sec:proof_privacy-recovery}
The scheme inherently satisfies the recovery and privacy conditions stated in Section~\ref{sec:rep-PMCsystem-model}. Privacy is guaranteed by satisfying the index, message, and database symmetry principles as for the PLC schemes in \cite{SunJafar19_2,ObeadKliewer18_1,ObeadLinRosnesKliewer18_1}. As for the recovery, one can easily see from the PIR query structure that the user is able to obtain all $\beta$ segments of the desired function based on the answers received from the $n$ databases. 
 Then, each segment is decoded (or optimally decompressed) to obtain in total $\beta\const{L}$ symbols with a probability of decoding error that is arbitrarily close to zero for a sufficiently large $\const{L}$.

\subsubsection{Achievable Rate}
\label{sec:proof_PCachievable-rate}

The PC rate of the scheme, assuming $\const{L} \to \infty$, is given by
\begin{IEEEeqnarray}{rCl}  
  \IEEEeqnarraymulticol{3}{l}{%
    \const{R}\overset{(a)}{=}\frac{\beta\const{L}\HH_\textnormal{min}}{\const{D}}
  }\nonumber\\
  & = &\frac{n^\mu\const{L}\HH_\textnormal{min}}{n\const{L}\biggl[\eHP{X^{[\mu]}}+\sum\limits_{\tau=2}^{\mu}{(n-1)}^{\tau-1}
    \sum\limits_{v=1}^{\mu-(\tau-1)}\binom{\mu-v}{\tau-1}\eHP{X^{(v)}}\biggr]}
  \nonumber\\
  & = &\frac{n^{\mu}\HH_\textnormal{min}}{n\biggl[\eHP{X^{[\mu]}}+\sum\limits_{\tau=2}^{\mu}{(n-1)}^{\tau-1}
    \sum\limits_{v=1}^{\mu-(\tau-1)}\binom{\mu-v}{\tau-1}\eHP{X^{(v)}}\biggr]}
  \IEEEeqnarraynumspace\label{eq:achievable-rate_PC}\\
  & \overset{(b)}{=} &\frac{n^{\mu-1}\HH_\textnormal{min}}{\eHP{X^{[\mu]}}
    +\sum\limits_{v=1}^{\mu-1}\sum\limits_{\tau=2}^{\mu-(v-1)}(n-1)^{\tau-1}\binom{\mu-v}{\tau-1}\eHP{X^{(v)}}}
  \nonumber\\
  & \overset{(c)}{=} &\frac{n^{\mu-1}\HH_\textnormal{min}}{\eHP{X^{[\mu]}}
    +\sum\limits_{v=1}^{\mu-1}\eHP{X^{(v)}}\sum\limits_{\tau'=1}^{\mu-v}\binom{\mu-v}{\tau'}(n-1)^{\tau'}}
  \nonumber\\
  & \overset{(d)}{=} &\frac{n^{\mu-1}\HH_\textnormal{min}}{\eHP{X^{[\mu]}}
    +\sum\limits_{v=1}^{\mu-1}\eHP{X^{(v)}}(n^{\mu-v}-1)}
  \nonumber\\
  & = &\frac{\HH_\textnormal{min}}{\sum\limits_{v=1}^{\mu-1}\frac{1}{n^{v-1}}\eHP{X^{(v)}}
    +\frac{1}{n^{\mu-1}}\Bigl[\eHP{X^{[\mu]}}-\sum\limits_{v=1}^{\mu-1}\eHP{X^{(v)}}\Bigr]},\nonumber
\end{IEEEeqnarray}
where $(a)$ follows from Definition~\ref{def:def_PCrate}, $(b)$ follows from changing the order of the two
summations, $(c)$ results by defining $\tau'=\tau-1$ of the second summation term, and $(d)$ follows from the binomial identity.

For the scenario of Corollary~\ref{cor:R_PCbasis}, by a similar approach of redundancy elimination as in the schemes in \cite[Sec.~III-C]{ObeadLinRosnesKliewer19_1app}, the PC scheme above can be modified by removing the redundant $1$-sums. %
Using \cite[Lem.~1]{ObeadLinRosnesKliewer19_1app} and $\eHP{X^{(v)}}=\HH_\textnormal{max}=1$, $\forall\,v\in[f]$, the PC rate can be shown to be equal to \eqref{eq:R_PCbasis}. %

%%%%%%%%%%%%%%%%%%%%%%%%%%%%%%%%%%%%%%%%%%%%%%%%%%%%%%%%%%%%%%%%%%%%%%%%%%%%%%%%%%%%%%%%%%%%%%%%%%%%%%%%%%%%%%%%%%%%%%%%%
\section{Discussion of the Outer Bound of Theorem~\ref{th:converse_bound}}
\label{sec:discussion_upper-bound}
By expanding the denominator of \eqref{eq:R_upperbound}, denoted by $\const{D}_\textnormal{opt}$, we get
\begin{IEEEeqnarray*}{rCl}
  \const{D}_\textnormal{opt} &=&\sum\limits_{v=1}^\mu n^{\mu-v+1}\bigl[\eHP{X^{[v]}}-\eHP{X^{[v-1]}}\bigr]\nonumber\\ %
  & = &
  n\bigHP{X^{[\mu]}}+n(n-1)\bigHP{X^{[\mu-1]}}\nonumber\\
  && +\> n(n-1)\cdot n\bigHP{X^{[\mu-2]}}+\cdots\nonumber\\
  && +\> n(n-1)\cdot n^{\mu-2}\eHP{X^{(1)}}.
\end{IEEEeqnarray*}

Next, consider the total download cost of the achievable scheme for Theorem~\ref{thm:PCachievable-rate} divided by
$\const{L}$, i.e., the denominator of \eqref{eq:achievable-rate_PC}, and denote it by $\const{D}_1$.  We have
\begin{IEEEeqnarray*}{rCl}
  \const{D}_1& = &n\eHP{X^{[\mu]}}+\sum_{\tau=2}^{\mu}n(n-1)^{\tau-1}
  \sum_{v=1}^{\mu-(\tau-1)}{\textstyle\binom{\mu-v}{\tau-1}}\eHP{X^{(v)}}
  \\
  & = &n\eHP{X^{[\mu]}}+n(n-1)\sum_{v=1}^{\mu-1}{\textstyle\binom{\mu-v}{1}}\eHP{X^{(v)}}\nonumber\\
  && +\> n(n-1)\sum_{v=1}^{\mu-2}(n-1){\textstyle\binom{\mu-v}{2}}\eHP{X^{(v)}}+\cdots\nonumber\\
  && +\> n(n-1)\cdot (n-1)^{\mu-2}\eHP{X^{(1)}}.
\end{IEEEeqnarray*}
By comparing $\const{D}_\textnormal{opt}$ with $\const{D}_1$, it can be seen that because joint compression
of the virtual message segments is not utilized, the outer bound of Theorem~\ref{th:converse_bound} is not achieved. An open question is to design an optimal scheme that achieves a download cost of $\const{D}_\textnormal{opt}$. 

%%%%%%%%%%%%%%%%%%%%%%%%%%%%%%%%%%%%%%%%%%%%%%%%%%%%%%%%%%%%%%%%%%%%%%%%%%%%%%%%%%%%%%%%%%%%%%%%%%%%%%%%%%%%%%%%%%%%%%%%%
\section{Special Case: Private Monomial Computation}
\label{sec:PMC_Redundancy}
In this section, we consider the special case of PMC. 
One can easily see that the assumption of Corollary~\ref{cor:R_PCbasis} covers the scenario of PMC, which includes the $f$ independent messages as candidate functions. Hence, as $\const{L} \to \infty$, the rate in \eqref{eq:R_PCbasis} is achievable for PMC.

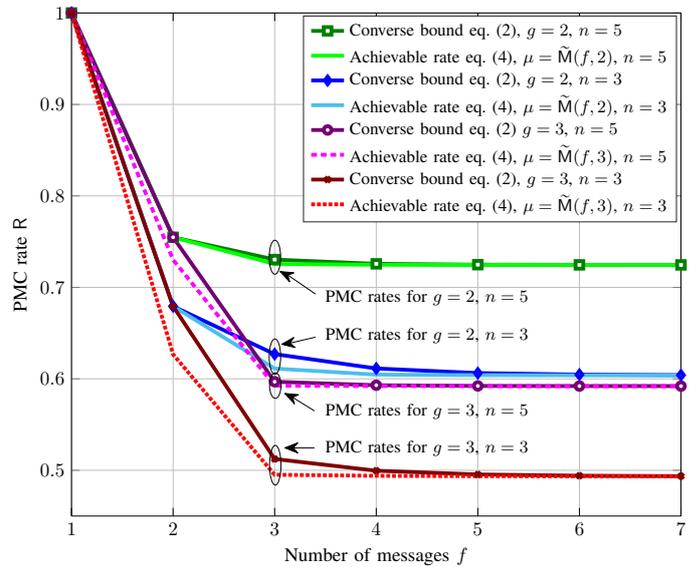
\begin{figure}[tbp!]
 	\centering
 	\resizebox{0.62\textwidth}{!} 
 	{% This file was created by matlab2tikz.
\definecolor{mycolor1}{rgb}{0.30196,0.74510,0.93333}% light blue
\definecolor{mycolor2}{rgb}{0.00000,0.49804,0.00000}% light green
%\definecolor{mycolor3}{rgb}{0.46667,0.67451,0.18824}% darker green
\definecolor{mycolor3}{rgb}{0.50196,0.00000,0.00000}% maroon
\definecolor{mycolor4}{rgb}{0.44902,0.00000,0.44902}% dark purble
\definecolor{mycolor5}{rgb}{1.00000,0.00000,1.00000}% magenta 
\begin{tikzpicture}

\begin{axis}[%
width=4.387in,
height=3.649in,
at={(0in,0.6in)},
scale only axis,
separate axis lines,
every outer x axis line/.append style={black},
every x tick label/.append style={font=\color{black}},
every x tick/.append style={black},
xmin=1,
xmax=7,
xtick={1, 2, 3, 4, 5, 6, 7},
xlabel={Number of messages $f$},
every outer y axis line/.append style={black},
every y tick label/.append style={font=\color{black}},
every y tick/.append style={gray},
ymin=0.45,
ymax=1,
ylabel={PMC rate $\mathsf{R}$},
axis background/.style={fill=white},
xmajorgrids,
ymajorgrids,
legend style={at={(0.38,0.584)}, anchor=south west, legend cell align=left, align=left, draw=white!15!black, fill=white} , cycle list
]
\addplot+[color=mycolor2, line width=2.0pt, mark=square*, mark options={mycolor2, fill=white}]
table[row sep=crcr]{%
	1	1\\
	2	0.754760498344864\\
	3	0.73041338549503\\
	4	0.725731248408523\\
	5	0.724802015055888\\
	6	0.724616453864117\\
	7	0.724579353026789\\
};
\addlegendentry{\small Converse bound eq.~\eqref{eq:PMCcapacity}, $g=2$, $n=5$}

\addplot [color=green, line width=2.0pt]
table[row sep=crcr]{%
	1	1\\
	2	0.754760498344864\\
	3	0.725664272806857\\
	4	0.724681083948884\\
	5	0.724591941658101\\
	6	0.724574450778674\\
	7	0.724570952880352\\
};
\addlegendentry{\small Achievable rate eq.~\eqref{eq:R_PCbasis}, $\mu=\widetilde{\mathsf{M}}(f,2)$, $n=5$}
\addplot[color=blue, line width=2.0pt, mark=diamond, mark options={solid, blue}]
  table[row sep=crcr]{%
1	1\\
2	0.679284448510378\\
3	0.627031798624964\\
4	0.61135600365934\\
5	0.606303474703478\\
6	0.60463780581693\\
7	0.604084614777756\\
};
\addlegendentry{\small Converse bound eq.~\eqref{eq:PMCcapacity}, $g=2$, $n=3$}

\addplot [color=mycolor1, line width=2.0pt]
  table[row sep=crcr]{%
1	1\\
2	0.679284448510378\\
3	0.611259007076291\\
4	0.604632779307829\\
5	0.604043576720553\\
6	0.603886505769681\\
7	0.603834431589487\\
};
\addlegendentry{\small Achievable rate  eq.~\eqref{eq:R_PCbasis}, $\mu=\widetilde{\mathsf{M}}(f,2)$, $n=3$}

\addplot [color=mycolor4, line width=2.0pt, mark=*, mark options={solid, fill=white}]
  table[row sep=crcr]{%
1	1\\
2	0.754760498344864\\ %0.616740451147394
3	0.596845597884574\\
4	0.593019664564802\\
5	0.592260356415551\\
6	0.592108728060094\\
7	0.592078411705168\\
};
\addlegendentry{\small Converse bound eq.~\eqref{eq:PMCcapacity} $g=3$, $n=5$}

\addplot [color=mycolor5, densely dashed, line width=2.0pt]
  table[row sep=crcr]{%
1	1\\
2	0.730074298724149\\ % 0.596568518561386
3	0.592517786828101\\
4	0.592160166291321\\
5	0.592088697583111\\
6	0.592074405911579\\
7	0.592071547660064\\
};
\addlegendentry{\small Achievable rate  eq.~\eqref{eq:R_PCbasis}, $\mu=\widetilde{\mathsf{M}}(f,3)$, $n=5$}

\addplot [color=mycolor3, line width=2.0pt, mark=x, mark options={solid,fill=white}]
  table[row sep=crcr]{%
1	1\\
2	0.679284448510378\\ %0.555066406032654\\
3	0.512368990183989\\
4	0.499559765429389\\
5	0.495431172326667\\
6	0.494070097677418\\
7	0.493618066481006\\
};
\addlegendentry{\small Converse bound eq.~\eqref{eq:PMCcapacity}, $g=3$, $n=3$}

\addplot [color=red, densely dotted, line width=2.0pt]
  table[row sep=crcr]{%
1	1\\
2	0.626725795844062\\  %0.512118944913257\\
3	0.495127314659448\\
4	0.493967359681366\\
5	0.49358387837287\\
6	0.493456183553774\\
7	0.493413633295413\\
};
\addlegendentry{\small Achievable rate  eq.~\eqref{eq:R_PCbasis}, $\mu=\widetilde{\mathsf{M}}(f,3)$, $n=3$}

\end{axis}

\begin{axis}[%
width=5.687in,
height=2.557in,
at={(0in,0.58in)},
scale only axis,
every outer x axis line/.append style={black},
every x tick label/.append style={font=\color{black}},
every x tick/.append style={black},
xmin=0,
xmax=1,
every outer y axis line/.append style={black},
every y tick label/.append style={font=\color{black}},
every y tick/.append style={black},
ymin=0,
ymax=0.5,
axis line style={draw=none},
ticks=none,
axis x line*=bottom,
axis y line*=left
]
\draw [black] (axis cs:0.257449,0.3710) ellipse [x radius=0.00725024, y radius=0.0241248]; % first ellipse from the top
\addplot [color=black, forget plot] % the arrow head
  table[row sep=crcr]{%
0.308774991964	0.318808316586951\\
0.274755136533271	0.344955366676743\\
};

\addplot[area legend, table/row sep=crcr, patch, fill=black, faceted color=black, forget plot, patch table={%
0	1	2\\
3	4	5\\
}]
table[row sep=crcr] {% the arrow body
x	y\\
0.276251714218468	0.335861021444936\\
0.264098039215686	0.350692307692308\\
0.273700113941641	0.345316463454655\\
0.264098039215686	0.350692307692308\\
0.281489209523535	0.3489825347469\\
0.273700113941641	0.345316463454655\\
};
\node[left, align=right]
at (axis cs:0.587,0.313) {\small PMC rates for $g=2$, $n=5$};

\draw [black] (axis cs:0.257449,0.230) ellipse [x radius=0.00725024, y radius=0.0251248];
\addplot [color=black, forget plot]
table[row sep=crcr]{%
	0.308774991964	0.257651592206817\\
	0.276555924167156	0.249191511376531\\
};
\addplot[area legend, table/row sep=crcr, patch, fill=black, faceted color=black, forget plot, patch table={%
	0	1	2\\
	3	4	5\\
}]
table[row sep=crcr] {%
	x	y\\
	0.283039198477206	0.24290016421409\\
	0.266029411764706	0.247518142235123\\
	0.276555924167156	0.249191511376531\\
	0.266029411764706	0.247518142235123\\
	0.28140889398282	0.257284948383565\\
	0.276555924167156	0.249191511376531\\
};

\node[left, align=right]
at (axis cs:0.587,0.26) {\small PMC rates for $g=2$, $n=3$};

\draw [black] (axis cs:0.25791,0.189) ellipse [x radius=0.00725024, y radius=0.0181248];
\addplot [color=black, forget plot]
table[row sep=crcr]{%
	0.3075	0.153056458635704\\
	0.279700113941641	0.172316463454655\\
};

\addplot[area legend, table/row sep=crcr, patch, fill=black, faceted color=black, forget plot, patch table={%
	0	1	2\\
	3	4	5\\
}]
table[row sep=crcr] {%
	x	y\\
	0.282251714218468	0.162861021444936\\
	0.270098039215686	0.177692307692308\\
	0.279700113941641	0.172316463454655\\
	0.270098039215686	0.177692307692308\\
	0.287489209523535	0.1759825347469\\
	0.279700113941641	0.172316463454655\\
};

\node[left, align=right]
at (axis cs:0.587,0.153) {\small PMC rates for $g=3$, $n=5$};
\draw [black] (axis cs:0.25791,0.076) ellipse [x radius=0.00725024, y radius=0.0281248];

\addplot [color=black, forget plot]
table[row sep=crcr]{%
	0.3075	0.10061\\
	0.280718954248366	0.10061\\
};

\addplot[area legend, table/row sep=crcr, patch, fill=black, faceted color=black, forget plot, patch table={%
	0	1	2\\
	3	4	5\\
}]
table[row sep=crcr] {%
	x	y\\
	0.286437908496732	0.093352685050798\\
	0.270098039215686	0.100609579100145\\
	0.280718954248366	0.100609579100145\\
	0.270098039215686	0.100609579100145\\
	0.286437908496732	0.107866473149492\\
	0.280718954248366	0.100609579100145\\
};
\node[left, align=right]
at (axis cs:0.587,0.100) {\small PMC rates for $g=3$, $n=3$};

\end{axis}
\end{tikzpicture}%
}
 	% \vspace{-2.5ex}
 	\caption{PMC rate $\const{R}$ versus the number of messages $f$ for the retrieval of nonparallel monomials over the field $\Field_3$.}
 	\label{fig:PMC}
 	% \vspace{-1.5ex}
 \end{figure}

In Fig.~\ref{fig:PMC}, for the field $\Field_3$ and $n=3$ and $5$, we plot the PMC rate computed from  \eqref{eq:R_PCbasis} and the outer bound from \eqref{eq:PMCcapacity} as a function of the number of messages $f$ for $\mu=\widetilde{\const{M}}(f,g)$ with $g=2$ and $g=3$, where  $\widetilde{\const{M}}(f,g)$ denotes  the number of \emph{nonparallel} monomials \cite[Sec.~III-E]{ObeadLinRosnesKliewer19_1app}.  Note that the PMC rate is close to the outer bound even for a small number of messages. As $f \to \infty$, it follows from Remark~\ref{rem:remark1} that the PMC rate approaches $\HH_\textnormal{min}(1-1/n)$.

%%%%%%%%%%%%%%%%%%%%%%%%%%%%%%%%%%%%%%%%%%%%%%%%%%%%%%%%%%%%%%%%%%%%%%%%%%%%%%%%%%%%%%%%%%%%%%%%%%%%%%%%%%%%%%%%%%%%%%%%%
\section{Conclusion}
\label{sec:conclusion}
We presented a novel PC scheme for noncolluding replicated databases
  and the scenario of nonlinear computation and showed that the resulting
  PC rate equals the PC capacity as the message size grows for the case when
  the candidate functions are the independent messages and one arbitrary nonlinear 
  function of these. Moreover, the PC rate approaches an outer
  bound on the PC capacity and thus becomes the capacity itself when the number of messages grows. Finally, we compared the outer bound and the achievable rate for the special case of PMC.

\bibliographystyle{IEEEtran}
\bibliography{./defshort1,./biblioHY}

\begin{thebibliography}{10}
\providecommand{\url}[1]{#1}
\csname url@samestyle\endcsname
\providecommand{\newblock}{\relax}
\providecommand{\bibinfo}[2]{#2}
\providecommand{\BIBentrySTDinterwordspacing}{\spaceskip=0pt\relax}
\providecommand{\BIBentryALTinterwordstretchfactor}{4}
\providecommand{\BIBentryALTinterwordspacing}{\spaceskip=\fontdimen2\font plus
\BIBentryALTinterwordstretchfactor\fontdimen3\font minus
  \fontdimen4\font\relax}
\providecommand{\BIBforeignlanguage}[2]{{%
\expandafter\ifx\csname l@#1\endcsname\relax
\typeout{** WARNING: IEEEtran.bst: No hyphenation pattern has been}%
\typeout{** loaded for the language `#1'. Using the pattern for}%
\typeout{** the default language instead.}%
\else
\language=\csname l@#1\endcsname
\fi
#2}}
\providecommand{\BIBdecl}{\relax}
\BIBdecl

\bibitem{ChorGoldreichKushilevitzSudan95_1}
B.~Chor, O.~Goldreich, E.~Kushilevitz, and M.~Sudan, ``Private information
  retrieval,'' in \emph{Proc. 36th Annu. IEEE Symp. Found. Comp. Sci. (FOCS)},
  Milwaukee, WI, USA, Oct. 23--25, 1995, pp. 41--50.

\bibitem{Gasarch04_1}
W.~Gasarch, ``A survey on private information retrieval,'' \emph{Bull. Eur.
  Assoc. Theor. Comput. Sci. (EATCS)}, vol.~82, pp. 72--107, Feb. 2004.

\bibitem{Yekhanin10_1}
S.~Yekhanin, ``Private information retrieval,'' \emph{Commun. ACM}, vol.~53,
  no.~4, pp. 68--73, Apr. 2010.

\bibitem{SunJafar17_1}
H.~Sun and S.~A. Jafar, ``The capacity of private information retrieval,''
  \emph{IEEE Trans. Inf. Theory}, vol.~63, no.~7, pp. 4075--4088, Jul. 2017.

\bibitem{SunJafar18_2}
------, ``The capacity of robust private information retrieval with colluding
  databases,'' \emph{IEEE Trans. Inf. Theory}, vol.~64, no.~4, pp. 2361--2370,
  Apr. 2018.

\bibitem{SunJafar19_2}
------, ``The capacity of private computation,'' \emph{IEEE Trans. Inf.
  Theory}, vol.~65, no.~6, pp. 3880--3897, Jun. 2019.

\bibitem{MirmohseniMaddahAli18_1}
M.~Mirmohseni and M.~A. Maddah-Ali, ``Private function retrieval,'' in
  \emph{Proc. Iran Workshop Commun. Inf. Theory (IWCIT)}, Tehran, Iran, Apr.
  25--26, 2018, pp. 1--6.

\bibitem{ChenWangJafar18_1}
Z.~Chen, Z.~Wang, and S.~Jafar, ``The asymptotic capacity of private search,''
  in \emph{Proc. IEEE Int. Symp. Inf. Theory (ISIT)}, Vail, CO, USA, Jun.
  17--22, 2018, pp. 2122--2126.

\bibitem{ObeadKliewer18_1}
S.~A. Obead and J.~Kliewer, ``Achievable rate of private function retrieval
  from {MDS} coded databases,'' in \emph{Proc. IEEE Int. Symp. Inf. Theory
  (ISIT)}, Vail, CO, USA, Jun. 17--22, 2018, pp. 2117--2121.

\bibitem{ObeadLinRosnesKliewer18_1}
S.~A. Obead, H.-Y. Lin, E.~Rosnes, and J.~Kliewer, ``Capacity of private linear
  computation for coded databases,'' in \emph{Proc. 56th Allerton Conf.
  Commun., Control, Comput.}, Monticello, IL, USA, Oct. 2--5, 2018.

\bibitem{Karpuk18_1}
D.~Karpuk, ``Private computation of systematically encoded data with colluding
  servers,'' in \emph{Proc. IEEE Int. Symp. Inf. Theory (ISIT)}, Vail, CO, USA,
  Jun. 17--22, 2018, pp. 2112--2116.

\bibitem{RavivKarpuk19_1app}
N.~Raviv and D.~A. Karpuk, ``Private polynomial computation from {L}agrange
  encoding,'' in \emph{Proc. IEEE Int. Symp. Inf. Theory (ISIT)}, Paris,
  France, Jul. 7--12, 2019.

\bibitem{ObeadLinRosnesKliewer19_1app}
S.~A. Obead, H.-Y. Lin, E.~Rosnes, and J.~Kliewer, ``Private polynomial
  computation for noncolluding coded databases,'' in \emph{Proc. IEEE Int.
  Symp. Inf. Theory (ISIT)}, Paris, France, Jul. 7--12, 2019.

\end{thebibliography}

\end{document}